\documentclass[journal]{IEEEtran}

\usepackage{cite}
\usepackage{amsmath,amssymb,amsfonts}
\usepackage{graphicx, float, booktabs, multirow}
\usepackage{multirow, makecell}
\usepackage{textcomp,enumitem}
\usepackage{xcolor}
\usepackage[ colorlinks = true,
linkcolor = blue,
urlcolor = blue,
citecolor = red,
anchorcolor = green,
]{hyperref}
\def\BibTeX{{\rm B\kern-.05em{\sc i\kern-.025em b}\kern-.08em
    T\kern-.1667em\lower.7ex\hbox{E}\kern-.125emX}}

\usepackage{amsthm}

\usepackage{algorithm}
\usepackage[noend]{algpseudocode}

\begin{document}

\title{
Diversifying Moving Target Defense Against Adaptive Attacks Through Factorized Diffusion
}

\author{
Chenhan Xiao,~\IEEEmembership{Student Member,~IEEE,}
and~Yang~Weng,~\IEEEmembership{Senior Member,~IEEE.}
\thanks{
This work has been submitted to the IEEE for possible publication. Copyright may be transferred without notice, after which this version may no longer be accessible.}
\vspace{-10mm}
}
\maketitle

\begin{abstract}
Proactive defense offers a means to enhance power-system cybersecurity through controlled modifications of system configurations. One example is moving target defense (MTD), which uses distributed flexible AC transmission system (D-FACTS) devices to perturb transmission-line reactances against stealthy false data injection attacks. Existing MTD methods primarily rely on deterministic optimization, producing a single perturbation plan for each operating condition. However, repeated or highly correlated plans may expose predictable system configurations that attackers can learn and exploit. To address this limitation, this paper reformulates MTD as a conditional generative problem that produces diverse, secure, and physically feasible perturbation candidates. Learning such a distribution is challenging because MTD decisions jointly involve selecting branches, determining perturbation magnitudes, and coordinating adjustments across the network. To make this problem tractable, we factorize the decision distribution into three sequential conditional stages: branch selection, magnitude allocation, and continuous refinement. Graph-conditioned diffusion models are employed to capture the discrete and continuous decisions while accounting for network dependencies. Furthermore, a security–operation objective is incorporated into diffusion training to favor candidates with desirable defense effectiveness and operational costs.
Numerical studies on the IEEE 118- and 300-bus systems show that the proposed framework achieves AC feasibility comparable to that of optimization-based MTD baselines and higher security–operation values, while reducing online computation time by an order of magnitude. The generated candidate pool remains effective under changing operating conditions and device availability, providing diverse high-value alternatives for flexible online cyber defense.
\end{abstract}


\section{Introduction}

Modern power systems increasingly rely on cloud-based monitoring, third-party analytics, and data-driven tools to support situational awareness and operational control \cite{bidgely, verdigris}. This growing reliance on networked operational data also increases the exposure of power-system measurements to cyber manipulation \cite{mantravadi2020securing, stouffer2023guide, xiao2026st}. An attacker may know the power-system model \cite{liu2011false} or can infer it from historical operating data \cite{xiao2025}. They can craft a false data injection attack (FDIA) whose injected measurements are consistent with the assumed grid model. Such stealthy attacks is a classic problem which evade residual-based bad-data detection mechanisms \cite{liu2011false, 8686241, 8468098}.

\begin{figure*}[t]
    \centering
    \includegraphics[width=1\linewidth]{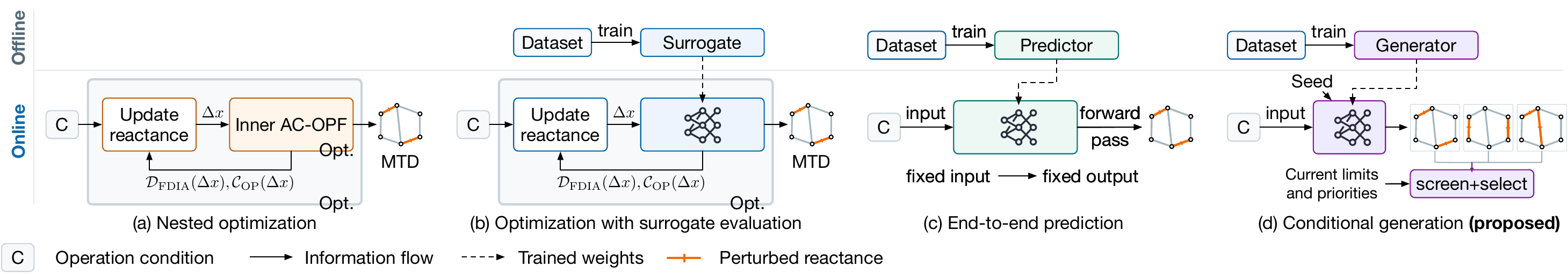}
    \caption{Comparison of D-FACTS-based MTD workflows.
    (a) Nested optimization repeatedly solves AC-OPF to evaluate the operational cost of candidate perturbations.
    (b) Surrogate-assisted optimization accelerates candidate evaluation while retaining the outer optimization loop.
    (c) End-to-end prediction maps each operating condition to a single MTD action.
    (d) The proposed conditional generator produces diverse MTD alternatives for each operating condition.
    }
    \label{fig:MTD-review}
    \vskip -0.1in
\end{figure*}

To counter stealthy FDIAs, proactive defense is promising as it leverage the advantage of creating a powerful asymmetric information contrast between the attacker and defender. For example, moving target defense (MTD) changes physical parameters to damage an attacker’s knowledge of grid model \cite{SubhashCost-Benefit}.
In transmission systems, moving target defense (MTD) can be implemented by adjusting line reactances through distributed flexible AC transmission system (D-FACTS) devices~\cite{9740216,9475527}. Although primarily used for power-flow regulation, D-FACTS devices can also support supplementary functions, such as oscillation damping and stability enhancement, without compromising their primary control duties~\cite{khalilian2011powerflow,khalilian2012ssr,mokhtari2013interaction}. This multifunctional capability motivates leveraging existing D-FACTS devices for MTD as an additional cybersecurity function~\cite{9018287}.
And, prior analysis suggests that hourly MTD updates may be practical given the time required for attackers to learn the perturbed measurement model~\cite{SubhashCost-Benefit}. Such updates can be coordinated with existing power-flow-control duties to limit additional switching and device wear.

Early D-FACTS-based MTD methods formulated MTD as an optimization problem. Optimal line-reactance perturbations are selected to maximize FDIA detectability (security benefit) while minimizing the associated operational cost, typically through mixed-integer optimization \cite{9740216,9475527}.
Subsequent studies extended this framework to time-varying designs \cite{10053623}, robustness analysis \cite{9905637}, and larger systems \cite{9018287,9497752,9966327,8416478}. 
Despite their physical grounding and demonstrated effectiveness, optimization-based methods can impose substantial online computational burdens \cite{XiaoSmartGridComm}. When operational cost is evaluated using AC-OPF, the outer MTD search repeatedly solves an inner AC-OPF problem to assess candidate reactance perturbations, resulting in a nested optimization structure, as shown in Fig.~\ref{fig:MTD-review} (a).
To reduce the cost of evaluating candidate MTD actions within the optimization loop, more studies have adopted surrogate metrics and models for the calculation of FDIA detectability and operational cost. For example, KL-divergence-based metrics quantify the distinguishability between pre- and post-perturbation models as an indicator of FDIA detectability \cite{mohammadpourfard2025adaptive}, while pre-trained neural-network surrogates estimate the operational cost induced by line-reactance perturbations \cite{liu2022explicit}. These techniques reduce the cost per evaluation but generally retain an outer optimization loop for selecting the MTD action, as shown in Fig.~\ref{fig:MTD-review} (b). 
More recent learning-based methods further accelerate online deployment by shifting the optimization burden to offline training. These methods learn a direct mapping from operating conditions to MTD decisions, so that online implementation only requires a forward pass of the trained model \cite{XiaoSmartGridComm, abdi2026proactive}, as illustrated in Fig.~\ref{fig:MTD-review} (c).

Despite their better consideration of different sencarios, existing MTD methods mostly use a deterministic design: a single reactance-perturbation plan is selected for each operating condition. This creates two limitations. First, from a security perspective, similar operating conditions may lead to the repeated deployment of identical or closely related reactance patterns, making the post-disturbance measurement model easier to infer. An attacker with partial or delayed knowledge of the deployed configuration could use observations across these deployments to learn the defense and develop more effective attacks.
Second, from an operational perspective, a single perturbation plan is operationally inflexible. At a given time, some D-FACTS devices may be assigned to power-flow control, making them unavailable for the prescribed MTD plan. Such conflicts may become more likely during contingencies \cite{10053623,mohammadpourfard2025adaptive}. The optimization can be re-solved with updated constraints, but this response is reactive and adds online computation burden.

An important observation is that a given operating condition does not necessarily have a unique MTD action with a favorable security–operation tradeoff. Due to the nonconvex nature of MTD design, multiple perturbation plans may achieve comparable security and operational performance while differing in their device utilization and reactance adjustments~\cite{liu2022explicit}. This observation motivates a shift from identifying a single optimized action to learning a distribution of diverse, high-quality MTD alternatives, as illustrated in Fig.~\ref{fig:MTD-review} (d). To capture this distribution, we construct an offline training dataset by repeatedly solving the optimization-based MTD formulation~\cite{SubhashCost-Benefit} from randomized initializations, collecting diverse feasible perturbation plans for each operating condition. Each action is further labeled with its FDIA detectability and incremental operational cost to guide distribution learning toward favorable security–operation tradeoffs. The resulting dataset combines optimization-informed solution quality with diverse perturbation patterns, enabling the generative model to learn multiple high-value MTD alternatives for online deployment.
At deployment, the trained model rapidly generates multiple MTD candidates with favorable security-operation tradeoffs for the current operating condition. These candidates can be screened against current system and device constraints to provide feasible alternatives without rerunning the full optimization for each set of operator priorities. Moreover, operators can vary the deployed defense across similar operating conditions, providing a mechanism to reduce repetition that can be exploited by an adaptive attacker. 

However, learning a distribution of high-quality MTD actions is challenging because each action involves multiple interdependent decisions. Specifically, the defender must determine which transmission lines to perturb, how much reactance adjustment to apply to each selected line, and how to coordinate these adjustments across the network to achieve favorable security–operation tradeoffs. These decisions are closely coupled through the network physics, making their joint distribution difficult to capture directly. To address this challenge, we factorize MTD generation into three sequential stages, motivated by~\cite{he2026powergrow}: branch selection, perturbation magnitude allocation, and continuous refinement. The first stage identifies the branches to be perturbed through binary decisions, the second determines their initial reactance adjustments through continuous decisions, and the third jointly refines the resulting perturbations to improve their security–operation performance. This decomposition follows the natural process of MTD design, allowing each stage to learn its decisions conditioned on those made previously rather than modeling all decisions simultaneously. To capture these conditional distributions, we employ graph-conditioned diffusion models~\cite{GBD,zhou2023beta}, which account for transmission-network dependencies and accommodate both discrete and continuous decision variables.

Another challenge is to direct candidate generation toward MTD actions with strong FDIA detectability and small operational cost. Standard diffusion training reproduces the empirical action distribution without explicitly prioritizing such security-operation value \cite{he2026powergrow, xiao2026operationally}. Consequently, low-value actions may dominate the candidate pool, increasing operators’ screening workload and limiting the availability of useful alternatives. To address this, we adopt reward-weighted diffusion training, using each reference MTD action’s security-operation value to weight the losses consistently across all three generation stages. Errors on higher-valued actions receive greater penalties, encouraging the models to generate candidates with more favorable security-operation tradeoffs.

Numerical studies on the IEEE 118- and 300-bus systems demonstrate that the proposed framework achieves favorable security–operation performance with substantially reduced online computation time compared with existing MTD methods. The results further highlight the value of generative MTD in providing diverse, high-quality defense alternatives that accommodate changing device availability and reduce attacker effectiveness through defense diversification. 
The paper is organized as follows. Section~\ref{sec:modeling} defines the measure of FDIA detectability and operational cost, and formulates the conditional generation problem.
Section~\ref{sec:method} develops a three-stage conditional generative framework that factorizes MTD action generation into branch selection, perturbation magnitude allocation, and continuous refinement, with each stage implemented by a graph diffusion model.
Section~\ref{sec:numerical} presents numerical studies. Section~\ref{sec:conclusion} concludes the paper.

\section{System Modeling}
\label{sec:modeling}

This section defines the physical constraints, FDIA detectability measure, and operational cost of D-FACTS-based MTD actions. Then it formulates the learning problem for generating high-quality candidate MTD actions conditioned on the current operating scenario.

Consider an AC transmission network represented by the graph $\mathcal{G}=(\mathcal{V},\mathcal{E})$, with $n$ buses, $m$ branches, and $n_g$ generators.
Let $x\in\mathbb{R}^{m}$ denote the branch reactances at the current device setpoints. An MTD action $\Delta x\in\mathbb{R}^{m}$ modifies selected branch reactances through D-FACTS devices, yielding the post-MTD reactance vector $x+\Delta x$.  Network connectivity and all other grid parameters, including line resistances, remain unchanged during the MTD process.
The MTD actions are evaluated according to two criteria \cite{SubhashCost-Benefit, XiaoSmartGridComm}. First, device admissibility requires $\Delta x$ to respect D-FACTS availability and adjustment limits. Then, the quality of the MTD action is reflected by the tradeoff between improved FDIA detectability and the associated operational cost. We formalize these criteria below.


\textbf{D-FACTS device admissibility.}
The reactance perturbation $\Delta x$ must respect the capabilities and availability of the D-FACTS devices. For each branch $\ell\in\mathcal{E}$, let $b_{\ell}\ge0$ denote the maximum permitted reactance adjustment and $\nu_{\ell}\in\{0,1\}$ indicate fixed device installation. Setting $\nu_{\ell}=0$ excludes branches without suitable devices and devices reserved for other control duties. The device-admissible action set is
\begin{equation}
    \mathcal{X}=
    \left\{
        \Delta x\in\mathbb{R}^{m}:
        |\Delta x_{\ell}|\le\nu_{\ell}b_{\ell},\
        \ell\in\mathcal{E}
    \right\},
    \label{eq:device_admissible_mtd}
\end{equation}
where we assume $0<b_{\ell}<x_{\ell}$ so that the post-MTD reactances remain positive. 

\textbf{FDIA detectability and operational cost.}
We quantify FDIA detectability by the average distance between measurements corrupted by attacks constructed using the pre-MTD model and measurements consistent with the post-MTD model. 
This distance captures the residual inconsistency that remains after state estimation and directly aligns with the classic residual-based bad data detection (BDD) \cite{liu2011false, xiao2025}.

Let $s=[v^\top,\varphi^\top]^\top$ denote the voltage state, where $v\in\mathbb{R}^{n}$ is voltage magnitude and $\varphi\in\mathbb{R}^{n}$ is phase angle. The AC measurement model is
$
    z=h_x(s)+\varepsilon,
    \label{eq:measurement_model}
$
where $h_x$ maps the state to the measurements $z$ and depends on the network topology and electrical parameters, including the reactances $x$. The measurement noise $\varepsilon$ has covariance matrix $R\succ0$.
Under an ideal stealthy attack \cite{liu2011false, xiao2025, mostafa2025, xiao2026cross}, the attacker constructs an additive manipulation to the original measurement $z$ as:
\begin{equation}
    z_a=z+a_x(\delta), \quad a_x(\delta)=h_x(s+\delta)-h_x(s).
    \label{eq:attack}
\end{equation}
Consequently, $z_a=h_x(s+\delta)+\varepsilon$, so the attacked measurements remain consistent with the pre-MTD model at the biased state $s+\delta$. The attack can therefore alter the estimated state while evading conventional residual-based BDD.

After an MTD action $\Delta x$, the measurement model becomes $h_{x+\Delta x}$. If the attacker applies the additive manipulation $a_x(\delta)$ constructed using the pre-MTD model, the attacked measurements may no longer be consistent with any state under the post-MTD model. The resulting residual may exceed the BDD threshold, revealing the attack.
To quantify this inconsistency, we define the FDIA detectability measure as 
\begin{equation}
    \begin{aligned}
        \mathcal{D}_{\mathrm{FDIA}}(\Delta x)
        =\mathbb{E}_{(s, \delta)}\Bigl[
        \min_{\tilde{s}}\,
        \bigl\|&h_{x+\Delta x}(s)+a_x(\delta)\\
        -&h_{x+\Delta x}(\tilde{s})
        \bigr\|_{R^{-1}}^2
        \Bigr],
    \label{eq:mtd_detectability}
    \end{aligned}
\end{equation}
where $\|u\|_{R^{-1}}^2:=u^\top R^{-1}u$ is the squared norm weighted by the inverse measurement-noise covariance.
The expectation is taken jointly over a prescribed distribution of plausible attack displacements $\delta$ and prescribed distribution of feasible operating states $s$.
Under the same measurement-noise conditions and BDD threshold, a larger residual distance in \eqref{eq:mtd_detectability} generally makes an attack more readily detectable. Maximizing \eqref{eq:mtd_detectability} therefore favors MTD actions that increase FDIA detectability.

To quantify the operational impact of the reactance perturbation $\Delta x$, we follow \cite{SubhashCost-Benefit, XiaoSmartGridComm} and evaluate the change in generation cost under permitted redispatch. The generation cost is evaluated using the AC optimal power flow (AC-OPF) value, denoted as $C_{\mathrm{AC}}(x; P^d, Q^d)$. Under the same grid operational limits, it is affected by the topology (thus the reactance $x$), and the active power demand $P^d\in\mathbb{R}^{n}$ and the reactive power demand $Q^d\in\mathbb{R}^{n}$.
Thus, the incremental operational cost induced by the MTD action $\Delta x$ is
\begin{equation}
    \mathcal{C}_{\mathrm{op}}(\Delta x)
    =C_{\mathrm{AC}}(x+\Delta x; P^d, Q^d)-C_{\mathrm{AC}}(x; P^d, Q^d),
    \label{eq:mtd_operating_cost}
\end{equation}
where the pre-MTD $C_{\mathrm{AC}}(x; P^d, Q^d)$ is set as the baseline cost. Minimizing \eqref{eq:mtd_operating_cost} therefore favors MTD actions with lower generation costs relative to the pre-MTD baseline.

For a fixed grid with topology $\mathcal{G}$ and D-FACTS device limits
$(\{b_\ell\}_{\ell\in\mathcal{E}},
\{\nu_\ell\}_{\ell\in\mathcal{E}})$, we represent the varying operating scenario by 
$c=(x,P^d,Q^d)$, which captures the current branch-reactance settings and the active- and reactive-power demand profiles.
For each operating scenario $c$, we combine FDIA detectability and operational cost into
\begin{equation}
    \mathcal{J}(\Delta x;c)
    =
    \mathcal{D}_{\mathrm{FDIA}}(\Delta x;c)
    -\lambda\cdot\mathcal{C}_{\mathrm{op}}(\Delta x;c),
    \label{eq:mtd_objective}
\end{equation}
where $\lambda>0$ specifies the security-operation trade-off and accounts for the different scales of the two terms.
For each operating scenario $c$, we construct the offline labeled dataset by solving the optimization-based MTD problem \cite{SubhashCost-Benefit} multiple times from randomized initializations. The resulting solutions are evaluated for security-operation value in \eqref{eq:mtd_objective}. After infeasible solutions are removed, the retained actions form
\(
\mathcal{D}_c
=
\left\{
\bigl(\Delta x_i,\mathcal{J}(\Delta x_i;c)\bigr)
\right\}_{i=1}^{N_c},
\)
where $N_c$ is the size of the dataset for scenario $c$.
Randomized initializations expose multiple locally optimal or near-optimal solutions while preserving the physical constraints and AC feasibility of the grid.
We formulate the conditional MTD generation problem as follows:
\begin{itemize}
    \item \textbf{Problem:}
    Learn to generate diverse, AC-feasible MTD candidates for screening against current system and device constraints and selection according to operational priorities.

    \item \textbf{Given:}
    Offline labeled datasets $\{\mathcal{D}_c\}$ collected across training scenarios.

    \item \textbf{Required:}
    A conditional generative model $p_\theta(\Delta x\mid c)$ that produces feasible MTD candidates $\Delta x\in \mathcal{X}$ with favorable values of
    $\mathcal{J}(\Delta x;c)$.
\end{itemize}


\section{Factorized Conditional Generation of MTD Actions}
\label{sec:method}

Directly learning the conditional distribution $p_\theta(\Delta x\mid c)$ formulated in Section~\ref{sec:modeling} is challenging because $\Delta x\in\mathbb{R}^{m}$ combines discrete branch-selection variables with continuous perturbation decisions. A monolithic generator must capture both structures and their dependencies simultaneously, making the distribution difficult to approximate. We therefore factorize the conditional MTD action distribution as follows.

\subsection{Factorization of the MTD Action Distribution}
\label{subsec:action_decomposition}

We organize MTD action generation into three stages: selecting the lines to perturb, allocating initial perturbations, and refining these perturbations to account for their coupled effects across the network.

\begin{figure}[h]
 \centering
 \includegraphics[width=\linewidth]{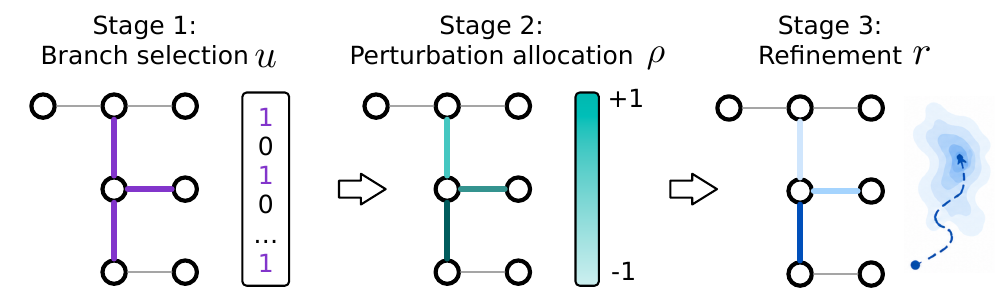}
 \caption{Three-stage factorization of MTD action generation.}
 \label{fig:factorization}
 \vskip -0.1in
\end{figure}

First, a binary vector $u\in\{0,1\}^{m}$ specifies the selected lines, where $u_\ell\le\nu_\ell$ restricts MTD to installed D-FACTS devices. Conditional on $u$, a normalized allocation vector $\rho\in[-1,1]^m$ specifies the initial perturbation magnitudes and directions, yielding
$\Delta x^{(0)}=b\odot u\odot\rho$, where $b=(b_\ell)_{\ell\in\mathcal{E}}$ and $\odot$ denotes elementwise multiplication. Finally, a correction vector $r\in[-r_{\max},r_{\max}]^m$, where $0<r_{\max}<1$, jointly refines the initial perturbations. The parameter $r_{\max}$ limits the refinement relative to the device adjustment bounds. This factorization process is shown in Fig. \ref{fig:factorization}. The resulting MTD action is
\begin{equation}
    \Delta x=b\odot u\odot
    \operatorname{clip}(\rho+r,-1,1),
    \label{eq:factorize}
\end{equation}
where clipping is applied elementwise. The joint distribution of these variables $(u,\rho,r)$ is modeled through the conditional factorization
\begin{equation}
        p_\theta(u,\rho,r\mid c)
        =p_{\theta_1}(u\mid c)\,
        p_{\theta_2}(\rho\mid c,u)p_{\theta_3}(r\mid c,u,\rho),
    \label{eq:factorized_distribution}
\end{equation}
where $\theta=(\theta_1,\theta_2,\theta_3)$.
Combined with \eqref{eq:factorize}, this joint distribution induces $p_\theta(\Delta x\mid c)$. The factorization makes the original complex distribution learning problem into three simpler learning tasks while retaining their dependencies through conditioning.

To construct training labels for the three stages, consider the $i$-th sample $(\Delta x_i,\mathcal{J}(\Delta x_i;c))$ in $\mathcal{D}_c$, where $i=1,\cdots,N_c$. We define the line-selection label as $u_{i,\ell}=\mathbf{1}{|\Delta x_{i,\ell}|\ge\tau}$, where $\tau>0$ is a prescribed threshold that identifies lines with appreciable perturbation magnitudes. Perturbations below this threshold are treated as inactive for line-selection labeling.
Then we extract the allocation vector as
$
    \rho_{i,\ell}^{\star}=
    u_{i,\ell}
        \Delta x_{i,\ell}/b_\ell.
$
To improve robustness to allocation errors, we deliberately introduce bounded variations around $\rho_i^\star$ and train the refinement stage to recover the reference allocation. Specifically, we construct the preliminary allocations and corresponding correction labels as
\begin{equation}
    \begin{aligned}
        \eta_i&\sim\textrm{Unif}([-r_{\max},r_{\max}]^m),\\
        \rho_i&=u_i\odot\operatorname{clip}
        (\rho_i^\star+\eta_i,-1,1),\\
        r_i&=\rho_i^\star-\rho_i,
    \end{aligned}
    \label{eq:allocation_residual_pairs}
\end{equation}
where $\textrm{Unif}$ denotes the uniform distribution. 
This construction ensures $r_i\in[-r_{\max},r_{\max}]^m$ and $\rho_i+r_i=\rho_i^\star$. Consequently, each augmented tuple $(u_i,\rho_i,r_i)$ reconstructs the original action $\Delta x_i$ through \eqref{eq:factorize} and retains its objective label $\mathcal{J}(\Delta x_i;c)$. These tuples provide training labels for the three conditional models in \eqref{eq:factorized_distribution}.

\subsection{Conditional Diffusion for the Three Generation Stages}
\label{subsec:diffusion_realization}

The three-stage factorization in \eqref{eq:factorized_distribution} provides a structured way to generate MTD actions, but each stage must still account for the physical characteristics of the transmission network. 
To accommodate these requirements, we implement the three conditional generation stages using Graph Beta Diffusion (GBD) models adapted from~\cite{GBD}. GBD combines graph-based representations with beta diffusion, allowing the same modeling framework to handle both continous and discrete variables within the MTD decisions.

All three models operate on the transmission network $\mathcal G$ and are conditioned on the current operating scenario $c=(x,P^d,Q^d)$. The D-FACTS installation indicators $\{\nu_\ell\}_{\ell\in\mathcal E}$ and adjustment limits $\{b_\ell\}_{\ell\in\mathcal E}$ are included in the branch attributes to distinguish controllable lines and their admissible perturbation ranges. The models also follow the dependency structure of the three-stage design: the allocation model $p_{\theta_2}(\rho\mid c,u)$ additionally receives $u$, while the refinement model $p_{\theta_3}(r\mid c,u,\rho)$ additionally receives both $u$ and $\rho$. 
Accordingly, for each training tuple $(u_i,\rho_i,r_i)$ constructed in Section~\ref{subsec:action_decomposition}, the conditioning information is given by
\begin{equation}
    \xi_1=c,\quad
    \xi_2=(c,u_i),\quad
    \xi_3=(c,u_i,\rho_i).
    \label{eq:stage_conditions}
\end{equation}
Following \cite{GBD}, we normalize the training targets to $u_i$, $(\rho_i+1)/2$, and $(r_i/r_{\max}+1)/2$ for branch selection, magnitude allocation, and refinement, respectively. Each normalized target is then mapped through a fixed affine transformation to an interval strictly inside $(0,1)$, ensuring positive beta-distribution parameters. These transformations are inverted after generation, and the branch-selection outputs are discretized.

For clarity, let $y$ denote the transformed target of any one stage, with the stage index omitted.
During training, beta diffusion progressively corrupts the reference decisions, enabling the model to learn how to recover the corresponding branch selection or reactance adjustments under the given operating condition. At diffusion step $t=1,\ldots,T$, the corrupted variable $y_t$ follows
\begin{equation}
    q_t(y_t\mid y)
    =
    \operatorname{Beta}\!\left(
        \kappa\alpha_t y,\,
        \kappa(1-\alpha_t y)
    \right),
    \label{eq:beta_forward_process}
\end{equation}
where $\kappa>0$ controls the concentration of the beta distribution and $\{\alpha_t\}_{t=0}^{T}$ is the decreasing diffusion schedule adopted from~\cite{GBD}. The distributions and arithmetic operations are applied coordinatewise. Unlike unrestricted Gaussian perturbations, the beta formulation keeps the diffused variables within $(0,1)$, consistent with the transformed ranges of the MTD decisions.

For stage $k\in\{1,2,3\}$, the graph-conditioned model estimates the reference decision from its corrupted representation and the available operating information: $\widehat y=\widehat y_{\theta_k}(y_t,t,\xi_k)$.
The predicted target remains within the same transformed interval and determines the reverse diffusion transition
\begin{equation}
    \begin{aligned}
        y_{t-1}
        &=y_t+(1-y_t)\odot\zeta_t,\\
        \zeta_t
        &\sim\operatorname{Beta}\!\left(
            \kappa(\alpha_{t-1}-\alpha_t)\widehat y,\,
            \kappa(1-\alpha_{t-1}\widehat y)
        \right).
    \end{aligned}
    \label{eq:beta_reverse_process}
\end{equation}
We denote the learned transition by $p_{\theta_k}(\cdot\mid y_t,t,\xi_k)$. When the predicted target $\widehat y$ is replaced by the reference target $y$, the transition becomes the corresponding forward-process posterior $q(\cdot\mid y_t,y)$. Repeated reverse transitions therefore allow each model to generate decisions consistent with its conditioning information, while different random samples can produce alternative MTD configurations.

To learn these transitions from the offline MTD solutions, each stage is trained to recover its reference decisions from their corrupted versions.
Following~\cite{GBD}, the training loss combines reverse-transition matching and marginal consistency:
\begin{equation}
    \begin{aligned}
        \ell_{k,i}
        ={}&(1-\omega)
        D_{\mathrm{KL}}\!\left(
            p_{\theta_k}(\cdot\mid y_t,t,\xi_k)
            \,\|\,q(\cdot\mid y_t,y)
        \right)\\
        &+\omega
        D_{\mathrm{KL}}\!\left(
            q_t(\cdot\mid\widehat y)
            \,\|\,q_t(\cdot\mid y)
        \right),
    \end{aligned}
    \label{eq:gbd_stage_loss}
\end{equation}
where $\omega\in[0,1]$ balances the two terms. The first term encourages the generated decisions to follow the reverse transitions associated with the reference MTD action, while the second promotes consistency between the distributions induced by the predicted and reference decisions. 

Learning from feasible MTD solutions alone does not ensure that the generated actions provide desirable defense performance. In particular, two feasible perturbation plans may differ substantially in their ability to expose FDIAs or in the operational cost incurred by their reactance adjustments.
If all training solutions are treated equally, the learned distribution primarily reflects how frequently different actions appear in the offline dataset, rather than their value for system defense.
As a result, the model may allocate a substantial portion of its generated candidates to actions with unfavorable security-operation tradeoffs.

To direct generation toward more effective MTD actions, we use the security-operation objective $\mathcal J(\Delta x;c)$ to guide distribution learning. Inspired by reward-guided diffusion training~\cite{DGPO}, we assign larger training weights to actions with higher security-operation values and minimize
\begin{equation}
    \mathcal{L}_{\mathrm{gen}}
    (\theta_1,\theta_2,\theta_3)
    =
    \mathbb{E}\!\left[
        e^{\beta\mathcal{J}(\Delta x_i;c)}
        \left(
            \ell_{1,i}+\ell_{2,i}+\ell_{3,i}
        \right)
    \right],
    \label{eq:weighted_generation_loss}
\end{equation}
where $\beta>0$ determines the emphasis placed on higher-valued MTD actions. The exponential weight increases the contribution of actions that offer more favorable security-operation tradeoffs, encouraging the models to reproduce such actions more accurately.
Importantly, the weight is assigned to the complete MTD action rather than to individual stage decisions. This is because the effectiveness of an MTD action depends on the combined branch selection and reactance adjustments, rather than on any single decision in isolation. Accordingly, the same weight is applied to all three stage losses and to all augmented training tuples reconstructed from the same reference action. This ensures that the three models are consistently guided toward decisions associated with favorable overall security–operation performance. The expectation in \eqref{eq:weighted_generation_loss} is taken over training scenarios, reference actions, allocation augmentation, diffusion times, and beta corruption. The three GBD models are then trained by minimizing the shared objective in \eqref{eq:weighted_generation_loss}, with each model updating its own parameters according to its corresponding weighted stage loss.

Once the three models are trained, a new operating scenario can be handled by sequentially generating the branch selection, magnitude allocation, and refinement decisions. For sampling run $j=1,\ldots,K_0$, the three decisions are drawn as
\begin{equation}
    \begin{aligned}
        u^{(j)}
        &\sim p_{\theta_1}(\cdot\mid c),\\
        \rho^{(j)}
        &\sim p_{\theta_2}(\cdot\mid c,u^{(j)}),\\
        r^{(j)}
        &\sim p_{\theta_3}
        (\cdot\mid c,u^{(j)},\rho^{(j)}),
    \end{aligned}
    \label{eq:sequential_candidate_sampling}
\end{equation}
using the reverse beta transitions in \eqref{eq:beta_reverse_process}.
The sampled decisions are mapped back to their physical ranges. Branch-selection outputs are discretized and masked by the D-FACTS installation indicators, while allocation and refinement outputs are restricted to their respective bounds. Perturbations on unselected branches are set to zero.
Finally, the three decisions are combined through \eqref{eq:factorize} to reconstruct reactance adjustments, forming a candidate pool $\{\Delta x^{(j)}\}_{j=1}^{K_0}$ of size $K_0$.

Because each sampling run has randomness, the framework can produce alternative branch combinations and reactance-adjustment patterns for the same operating condition. 
Training on feasible actions encourages the generation of AC-feasible candidates, with the resulting feasibility rate evaluated in Section~\ref{subsec:evaluate_AC}.
The weighted training objective further favors actions with higher security-operation values, as evaluated in Section~\ref{subsec:mtd_num_ablation}.
Moreover, the resulting candidate pool offers operators multiple MTD alternatives to be screened and deployed under similar operating conditions.
This variation can reduce repeated exposure of the same post-MTD network configuration and thereby make the defense more difficult for an adaptive attacker to learn, as investigated in Section~\ref{subsec:mtd_num_adaptive}. 
\section{Numerical Studies}
\label{sec:numerical}

The numerical studies evaluate whether the proposed framework
(1) generates MTD actions that admit feasible AC power-flow solutions and provide favorable security-operation values, 
(2) supplies diverse candidate plans that remain useful under changing D-FACTS restrictions,
(3) benefits from mixed-variable factorization, continuous refinement, and reward-weighted training, and
(4) maintains its generation quality and computational advantage as the network size increases.
In the following, the IEEE 118-bus system is used for the main comparisons and mechanism analysis, while the IEEE 300-bus system is used for the scalability study.

\subsection{Experimental Setup and Comparison Methods}
\label{subsec:mtd_num_setup}

For each system, we construct varying operating scenarios $c=(x,P^d,Q^d)$ by multiplying their nominal values by independently sampled factors from $\mathrm{Unif}(0.5,1.5)$. Network connectivity and all remaining electrical parameters are held fixed. We retain only scenarios for which MATPOWER \cite{zimmerman2010matpower} \texttt{runpf} converges. 
Approximately $20\%$ of branches are randomly selected~\cite{XiaoSmartGridComm}, with $\nu_\ell=1$ on these branches and zero elsewhere. This installation mask remains fixed across training and test datasets. The maximum permitted adjustment is $b_\ell=0.5x_\ell^{\mathrm{nom}}$, where $x_\ell^{\mathrm{nom}}$ is nominal physical reactance. 
For each scenario $c$, we repeatedly solve the MTD optimization problem \cite{SubhashCost-Benefit} from randomized initializations and retain $N_c=100$ feasible MTD solutions with distinct dominant device-use patterns. Two solutions are considered distinct if their six largest absolute perturbations occur on different sets of lines. We use six lines to characterize dominant device usage without distinguishing solutions solely by small perturbations on other lines.
Each retained action $\Delta x$ is labeled offline with its FDIA detectability $\mathcal{D}_{\mathrm{FDIA}}(\Delta x;c)$, incremental operational cost $\mathcal{C}_{\mathrm{op}}(\Delta x;c)$, and security-operation value $\mathcal J(\Delta x;c)$. 
To estimate the expectation in \eqref{eq:mtd_detectability}, we sample active and reactive demands by adding zero-mean Gaussian perturbations with standard deviations equal to $30\%$ of their nominal magnitudes, and solve the corresponding AC power flows in MATPOWER to obtain voltage states $s$. For each sampled state, we construct an attack perturbation $\delta=\mathcal N^{-1}(\mathcal N(s)+d)-s$, where $\mathcal N$ denotes min--max state normalization and $d\sim\mathrm{Unif}([0.1\mathbf1,\mathbf1])$ on free state coordinates, with the reference angle held fixed. We then evaluate the minimized residual in \eqref{eq:mtd_detectability} for each sampled state-attack pair and average these values to estimate the expectation.
For $\mathcal{C}_{\mathrm{op}}(\Delta x;c)$, the pre- and post-MTD operating costs in \eqref{eq:mtd_operating_cost} are computed using AC-OPF in MATPOWER.
We collect these scenario-specific datasets $\{\mathcal D_c\}$ for 1,000 operating scenarios and partition the scenarios into training and test sets with a 70\%/30\% split. The resulting training set contains $700\times100=70{,}000$ action-value pairs.  Table~\ref{tab:mtd_num_setup} summarizes the experimental configuration and model hyperparameters.

\begin{table}[t]
\centering
\caption{Experimental configuration and model hyperparameters.}
\label{tab:mtd_num_setup}
\small
\setlength{\tabcolsep}{3pt}
\renewcommand{\arraystretch}{1.08}
\begin{tabular}{p{0.56\columnwidth}p{0.36\columnwidth}}
\toprule
Parameter & Setting \\
\midrule
Scenario scaling factors
& $\mathrm{Unif}(0.5,1.5)$ \\

D-FACTS deployment fraction
& $20\%$ \\

Line-selection threshold
& $\tau_\ell = 0.05 x_\ell$ \\

Operating scenarios per test system
& $1{,}000$ \\

Retained actions per scenario $N_c$
& $100$ \\

Training/test split by scenario
& $70\%/30\%$ \\

\midrule
State construction $s$
& Gaussian demand perturtations plus AC power-flow solve\\

Attack-displacement distribution for $\delta$
& $\mathrm{Unif}([0.1\cdot\mathbf{1},\,\mathbf{1}])$ \\

Security-operation coefficient $\lambda$
& $10^{-5}$ \\

\midrule
Normalized refinement bound $r_{\max}$
& $0.5$ \\

Beta-diffusion concentration $\kappa$
& $30$ \\

Reverse-diffusion steps per stage $T$
& $1{,}000$ \\

Marginal-consistency loss weight $\omega$
& $0.01$ \\

Reward-weighting coefficient $\beta$
& $0.3$ \\

Nominal candidate-pool size $K_0$
& $5$ \\

Training epochs
& $3200$ \\
\bottomrule
\end{tabular}
\end{table}

We compare the proposed MTD framework with representative optimization-based,
learning-based, and generative-based MTD approaches.

\begin{itemize}[leftmargin=*]
\item \emph{Optimization-based MTD:}
We include the cost-benefit MTD method in~\cite{SubhashCost-Benefit}, which balances attack detectability and generation cost when selecting reactance perturbations and also serves as the basis for constructing our training dataset. We also consider robust MTD (RMTD)~\cite{9905637}, which optimizes the separation between pre- and post-MTD Jacobian subspaces, providing a benchmark for robustness to adverse attack directions. Finally, data-driven MTD (DD-MTD)~\cite{mohammadpourfard2025adaptive} combines a KL divergence criterion with Kolmogorov--Smirnov constraints through piecewise linear approximation and mixed-integer linear programming, providing a benchmark for computationally efficient optimization-based MTD.

\item \textit{Learning-based MTD.}
We consider neural-surrogate MTD (NS-MTD)~\cite{XiaoSmartGridComm}, which uses a residual-based detectability metric and a neural surrogate for operational cost, providing a benchmark for neural acceleration of optimization-based MTD. We also include TD3-based MTD~\cite{abdi2026proactive}, which learns continuous D-FACTS control through reinforcement learning, providing a policy-learning benchmark.

\item \textit{Alternative conditional generation.}
We include a conditional variational autoencoder (CVAE) adapted from~\cite{sohn2015learning}, which generates reactance perturbations on the same graph information and operating scenario. It is trained on the same offline dataset with comparable objective-dependent sample weighting, providing a benchmark for assessing the benefits of diffusion-based conditional generation.
\end{itemize}

We evaluate all methods on the test dataset using the following metrics: 
(1) the power-flow convergence rate based on MATPOWER \texttt{runpf}, representing the AC-feasibility rate,  
(2) the mean FDIA detectability $\bar{\mathcal{D}}_{\mathrm{FDIA}}$,
(3) the mean incremental operational cost $\bar{\mathcal{C}}_{\mathrm{OP}}$, (4) the mean security-operation value $\bar{\mathcal J}$,
(5) the empirical BDD detection rate for attacks constructed using the pre-MTD model according to~\eqref{eq:attack}, under a false-alarm probability $\alpha=5\%$, and 
(6) the online processing time per returned MTD action, excluding offline training.
Optimization-based and learning-based methods return one action per scenario, whereas generative methods produce $K_0$ candidates.
Power-flow convergence uses all produced MTD actions. The average of $\bar{\mathcal D}_{\mathrm{FDIA}}$, $\bar{\mathcal C}_{\mathrm{op}}$, and $\bar{\mathcal J}$ is calculated using the subset of AC-feasible MTD actions.

\begin{table*}[t]
\centering
\caption{Performance comparison of MTD methods on the IEEE 118-bus system. The BDD false-alarm rate is $\alpha=5\%$.}
\label{tab:mtd_num_main}
\small
\setlength{\tabcolsep}{7pt}
\begin{tabular}{lcccccc}
\toprule
Method &
\makecell{AC-feasibility rate \\ (\%)} &
\makecell{$\bar{\mathcal{D}}_{\mathrm{FDIA}}$ \\ \ } &
\makecell{$\bar{\mathcal{C}}_{\mathrm{OP}}$ \\ ($10^3\,\$/\mathrm{h}$)} &
\makecell{$\bar{\mathcal J}$ \\ \ } &
\makecell{BDD detection rate \\ (\%)} &
\makecell{Runtime per action \\ (s)} \\
\midrule
Cost-benefit MTD \cite{SubhashCost-Benefit} & 97.7   & 0.192 & 5.3 & 0.139 & 95.69 & 19.5 \\
Robust MTD \cite{9905637}  & 95.0 & 0.184 & 8.9 & 0.095 & 93.77 & 16.7 \\
Data-driven MTD \cite{mohammadpourfard2025adaptive} & 90.7  & 0.190 & 15.7 & 0.033 & 93.23 & 8.4 \\
\midrule
Neural-surrogate MTD \cite{XiaoSmartGridComm} & 77.3  & 0.201 & 12.9 & 0.072 & 94.43 & 0.78 \\
TD3-MTD \cite{abdi2026proactive} & 78.7  & 0.106 & 7.6 & 0.030 & 88.34 & 0.12 \\
\midrule
CVAE \cite{sohn2015learning} & 89.7  & 0.170 & 10.2 & 0.068 & 78.07 & 0.25 \\
Ours & 94.3 & 0.323 & 1.7 & 0.306 & 98.54 & 0.59 \\
\bottomrule
\end{tabular}
\end{table*}

All experiments are conducted on the ASU Sol high-performance computing cluster using four NVIDIA A100 GPUs with 80\,GB of memory each.
To stabilize diffusion training, input electrical parameters and demand profiles are normalized to $[0,1]$, and generated actions are mapped back to physical units. The diffusion models are trained using the Adam optimizer with a learning rate of $10^{-4}$ and a batch size of 50. Additional hyperparameters are reported in Table~\ref{tab:mtd_num_setup}.

\subsection{MTD Effectiveness and Computational Efficiency}
\label{subsec:evaluate_AC}

We first evaluate the overall effectiveness and computational efficiency of the proposed framework against representative MTD methods.
As shown in Table~\ref{tab:mtd_num_main}, optimization-based methods retain strong AC feasibility on the IEEE 118-bus system through explicit constraint enforcement, with only rare solver failures. In comparison, the proposed method achieves higher FDIA detectability and lower incremental operating cost, leading to improved $\bar{\mathcal J}$ and BDD detection rates while reducing online runtime by an order of magnitude.
This advantage arises from shifting substantial computation offline. Randomized initializations yield multiple locally optimized actions for each scenario, enabling the generator to learn favorable regions of the action space, while reward weighting further biases training toward higher-value solutions. This combination reduces the dependence on a single initialization in the nonconvex optimization of MTD problem, which may converge to a less favorable local solution. 

Compared with neural-surrogate and policy-based MTD, the proposed method achieves higher AC feasibility and a stronger security-operation tradeoff. NS-MTD maintains competitive detectability but incurs higher operational cost and more feasibility failures, whereas TD3-MTD offers fast inference with lower action quality. 
Although slower than direct policy inference, our approach maintains subsecond runtime per action with better security-operation performance.
The CVAE comparison further shows that conditional generation alone is insufficient: under the same offline dataset and comparable weighting, CVAE yields lower feasibility, security-operation value, and detection performance. These results support the proposed factorized diffusion framework for learning high-quality MTD action distributions, with component-wise contributions examined in the ablation study in Section \ref{subsec:mtd_num_ablation}.

\subsection{High-Quality MTD Alternatives for Flexible Deployment}
\label{subsec:mtd_num_alternatives}

We next examine whether the generated candidate pools provide multiple useful MTD alternatives at each operating point. We assess two complementary properties: the quality of individual candidates and the diversity of their dominant D-FACTS device usage patterns.

\begin{figure}[h]
    \centering
    \includegraphics[width=1\linewidth]{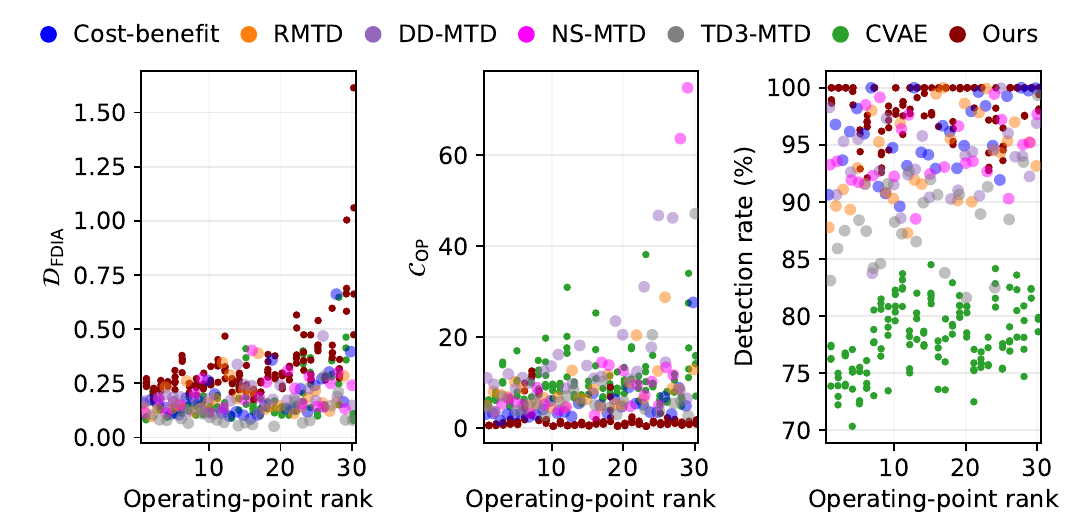}
    \caption{Per-operating-point $\mathcal{D}_{\mathrm{FDIA}}$, $\mathcal{C}_{\mathrm{op}}$ (in $10^3$), and BDD detection rate for 30 randomly selected IEEE 118-bus test scenarios. Operating points are ranked separately in each panel by the aggregate metric across methods. BDD false-alarm rate of $\alpha=5\%$.}
    \label{fig:mtd_num_quality}
\end{figure}

Fig.~\ref{fig:mtd_num_quality} randomly selects 30 scenarios from the 300-scenario IEEE 118-bus test set, and compares individual MTD action outcomes. It shows their corresponding $\mathcal{D}_{\mathrm{FDIA}}$, $\mathcal{C}_{\mathrm{op}}$ (in $10^3\,\$/\mathrm{h}$), and BDD detection rate from left to right. CVAE and the proposed method each generate $K_0=5$ candidates per operating point, whereas the other methods return one action. Operating points are ranked separately in each panel by the aggregate metric across methods, with candidate scores averaged within each generative method. 
In this comparison, the proposed method achieves higher FDIA detectability in general while maintaining low incremental costs and high BDD detection rates. Compared with another generative baseline CVAE, the proposed method shows stronger detection performance and less cost dispersion, providing multiple favorable alternatives in the displayed subset for operator selection.

\begin{table*}[t]
\centering
\caption{Ablation study on the IEEE 118-bus system. The BDD false-alarm rate is $\alpha=5\%$.}
\label{tab:mtd_num_ablation}
\small
\setlength{\tabcolsep}{6pt}
\begin{tabular}{lcccccc}
\toprule
Variant &
\makecell{AC-feasibility rate\\(\%)} &
$\bar{\mathcal{D}}_{\mathrm{FDIA}}$ &
\makecell{$\bar{\mathcal{C}}_{\mathrm{OP}}$\\($10^3\,\$/\mathrm{h}$)} &
$\bar{\mathcal J}$ &
\makecell{BDD detection rate\\(\%)} &
\makecell{Runtime per action\\(s)} \\
\midrule
Joint diffusion (one model)
& 85.9 & 0.263 & 3.6 & 0.227 & 94.82 & 0.54 \\
Without refinement
& 90.8 & 0.296 & 2.4 & 0.272 & 96.91 & 0.41 \\
Without reward weighting
& 94.5 & 0.312 & 2.0 & 0.292 & 98.07 & 0.59 \\
\midrule
\textbf{Full model}
& 94.3 & 0.323 & 1.7 & 0.306 & 98.54 & 0.59 \\
\bottomrule
\end{tabular}
\end{table*}

\begin{figure}[h]
    \centering
    \includegraphics[width=1\linewidth]{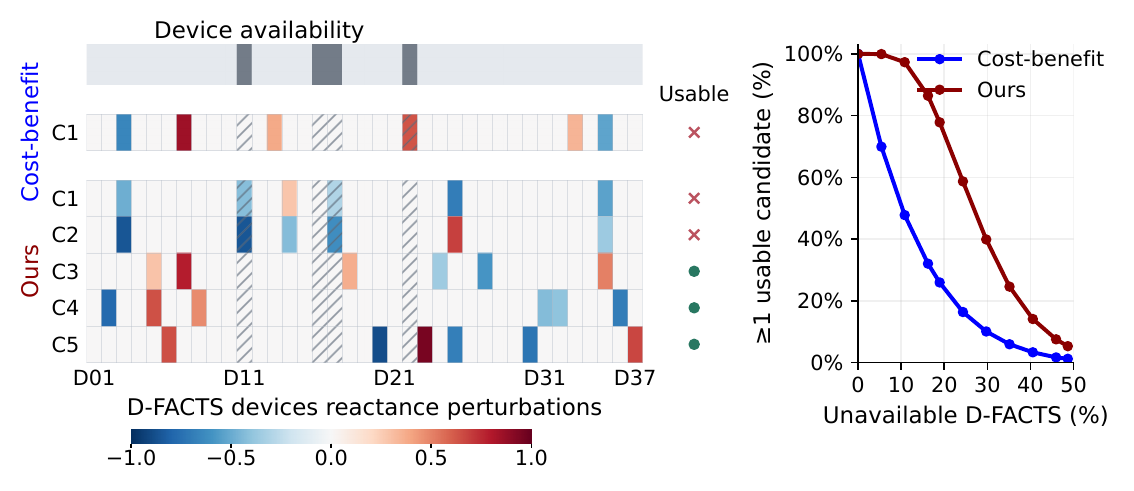}
    \caption{Left: the six largest normalized reactance perturbations for one cost-benefit MTD action and $K_0=5$ proposed candidates at the same operating point on the 118-bus system. Right: coverage of usable complete actions across 300 test scenarios and ten random masks per scenario at each unavailability level. Usability requires device compatibility and AC feasibility.}
    \label{fig:mtd_device_diversity}
\end{figure}

We further examine whether the generated MTD alternatives exhibit different dominant-device patterns and accommodate changing device availability. 
Alternative perturbation patterns allow operators to vary deployed perturbation patterns, potentially reducing their predictability to attackers, while respecting existing operating schedules and newly imposed device commitments.
The left panel of Fig.~\ref{fig:mtd_device_diversity} compares cost-benefit MTD with the proposed method at one operating point. 
For visualization, we display the six largest absolute perturbations and hide the remaining entries. All physical and security evaluations, including device compatibility, use the complete original actions.
Four of the 37 D-FACTS devices are randomly designated unavailable, corresponding to approximately $10\%$. 
The single cost-benefit action selected for this illustration conflicts with an unavailable device and therefore offers no compatible deployment option without recomputation. In contrast, although two of the proposed method's $K_0=5$ candidates encounter similar conflicts, the remaining three are classified as device-compatible and AC-feasible. Their different perturbation patterns provide alternative configurations under the same operating condition.

The single action returned by cost-benefit MTD conflicts with an unavailable device and therefore offers no compatible deployment option without recomputation. In contrast, although two of the proposed method's $K_0=5$ candidates encounter similar conflicts, the remaining three avoid the unavailable devices. Their different activation patterns provide alternative configurations under the same operating condition.

The right panel extends this illustrative comparison by evaluating all 300 test scenarios, varying the unavailable-device fraction from 0 to approximately $50\%$ and drawing ten random device masks per scenario at each level. The same masks are applied to both methods. Coverage is the proportion of scenario--mask pairs retaining at least one action that is both device-compatible and AC-feasible. The shown results indicate that the coverage decreases for both methods as restrictions increase, but the proposed candidate pool exhibits a substantially slower decline. At approximately $20\%$ device unavailability, it retains a compatible alternative for about $80\%$ of operating points, compared with roughly $25\%$ for cost-benefit MTD. This pattern shows that multiple distinct candidates can preserve deployment options when individual actions become unusable, allowing operators to respond to changing device commitments by screening an existing pool without resolving the MTD optimization problem.

\subsection{Resistance to Adaptive FDIAs via Defense Diversification}
\label{subsec:mtd_num_adaptive}

We evaluate whether variation in deployed MTD actions can limit an attacker's reuse of information learned from post-MTD measurements. 
We compare five deployment strategies: cost-benefit MTD, highest-value and random selection for CVAE-based MTD (\emph{CVAE: highest} and \emph{CVAE: random}), and the corresponding policies for the proposed framework (\emph{Ours: highest} and \emph{Ours: random}). For each generative method and scenario instance, the two policies share the same raw candidate pool of size $K_0=5$. After feasibility screening, the highest-value policy selects the candidate with the largest $\mathcal J$, whereas the random policy selects uniformly from the feasible candidates. 

To first examine MTD action variability under similar operating conditions, we select 30 reference scenarios from the IEEE 118-bus test set and generate nearby scenarios through small demand perturbations. For each reference scenario and each perturbation level from $0\%$ to $5\%$ (in increments of $0.5\%$), we generate 10 additional scenarios by adding zero-mean Gaussian noise to the active and reactive demands. For each demand component, the noise standard deviation is the specified percentage of the absolute value of its reference demand. This yields 300 additional scenario instances per perturbation level. Branch reactances and all other network parameters remain fixed.
We apply each method to all reference and perturbed operating scenarios, selecting one MTD action per scenario according to its deployment policy.
Each perturbed scenario is compared with its corresponding reference scenario. 
For the corresponding MTD actions $\Delta x$ and $\Delta x^{\mathrm{ref}}$, we define the mean squared normalized action distance as
\(
d_{\mathrm{act}}(\Delta x,\Delta x^{\mathrm{ref}})
=
\frac{1}{|\mathcal E_{\mathrm D}|}
\sum_{\ell\in\mathcal E_{\mathrm D}}
(
\frac{\Delta x_\ell-\Delta x_\ell^{\mathrm{ref}}}{b_\ell}
)^2,
\)
where $\mathcal E_{\mathrm D}$ is the set of branches equipped with D-FACTS devices, and $b_\ell$ is the adjustment bound for branch $\ell$. This calculation includes all entries of the original action vectors. At each perturbation level, we average both distances over the $30\times10=300$ reference--perturbation pairs. 

\begin{figure}[h]
    \centering
    \includegraphics[width=1\linewidth]{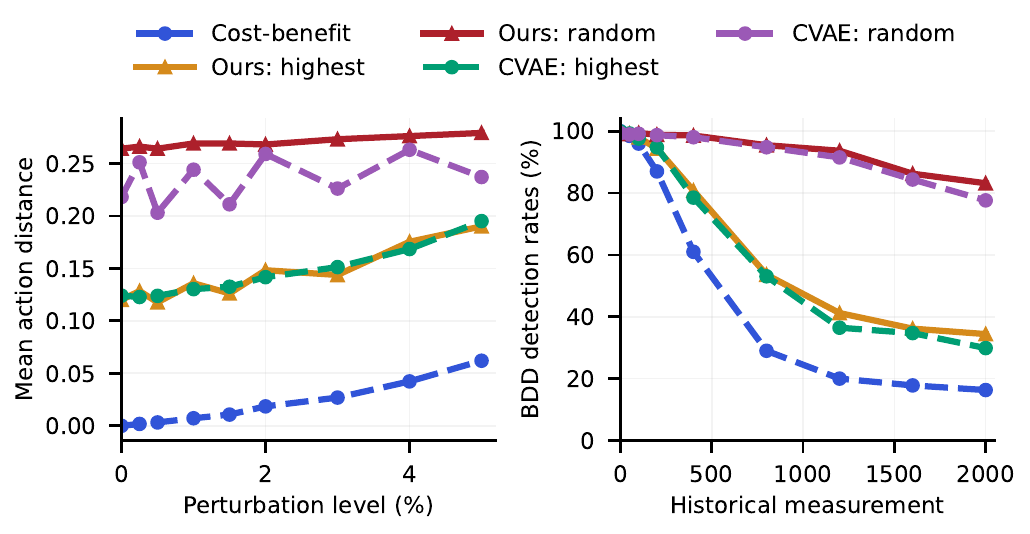}
    \caption{Left: Mean action distance from reference-scenario actions versus demand perturbation level. Right: BDD detection rate versus the number of historical measurement samples available to the attacker, with the operation scenario perturbation level fixed at $2\%$. }
    \label{fig:mtd_num_adaptive}
\end{figure}

The left panel of Fig.~\ref{fig:mtd_num_adaptive} plots the mean action distance against the perturbation level for each method.
It shows that cost-benefit MTD tends to select similar actions under nearby operating conditions. Both generators introduce greater action variation, although highest-value selection produces less separation than random selection. The proposed random-selection policy maintains the largest mean action distance even when operating conditions change little. This highlights the potential of generative MTD to reduce recurring defense patterns and thereby limit opportunities for attackers to reuse previously learned information.

We next evaluate whether the observed variation in MTD actions reduces attacker effectiveness. Fixing the operation scenario perturbation level at $2\%$, we apply each method's selected actions to the IEEE 118-bus system and solve the AC power flow in MATPOWER under the corresponding demand profiles. The resulting measurement sequences are concatenated to form the historical observations available to an attacker. Following the data-driven AC FDIA construction in~\cite{xiao2025}, the attacker trains an autoencoder on these observations to approximate the measurement manifold and construct attack measurements intended to evade residual-based BDD. For each observation budget, we generate 100 attack measurement vectors and report the BDD detection rate. By varying the number of historical measurement samples available to the attacker, this experiment assesses whether defense variation limits the attacker's ability to learn from accumulated observations and construct attacks that evade detection.

The right panel of Fig.~\ref{fig:mtd_num_adaptive} shows that BDD detection declines as the attacker accumulates historical measurements, with the sharpest decrease under cost-benefit MTD. Highest-value selection slows this decline for both generators, while random selection maintains substantially higher detection rates. Together with the action-variability comparison, these trends suggest that varying the deployed configuration limits the usefulness of accumulated observations for constructing evasive attacks. The proposed random-selection policy retains the highest detection rate at larger observation budgets, highlighting the potential security benefit of deploying diverse MTD alternatives.

\subsection{Ablation Study: Impact of Key Design Components}
\label{subsec:mtd_num_ablation}

To assess the contribution of each key design component to MTD performance, we conduct three ablation experiments that separately examine the effects of mixed-variable factorization, continuous refinement, and security–operation-weighted training.
To this end, we consider the following model variants:
(1) \emph{Joint diffusion} replaces the three conditional diffusion models with one model that jointly generates $(u,\rho,r)$, retaining the same action reconstruction and a comparable total parameter count.
(2) \emph{Without refinement} removes the third diffusion model, sets $r=0$, and trains the allocation model directly on the clean targets $\rho_i^\star$, without the allocation corruption used for refinement training.
(3) \emph{Without reward weighting} retains all three models but sets $\beta=0$ in~\eqref{eq:weighted_generation_loss}, assigning equal weights to training MTD actions.
All variants use the same training dataset and candidate budget $K_0=5$ during generation.

We report the averaged metrics over all operation points of these model variants in Table~\ref{tab:mtd_num_ablation}.
The joint model's lower AC feasibility and security-operation value support the benefit of explicitly separating branch selection from conditional perturbation generation. A single model must simultaneously recover discrete device choices and continuous settings, whereas factorization structures these dependencies into successive conditional learning tasks. The difference under comparable model capacity therefore suggests that the organization of generation contributes to action quality. Removing refinement illustrates a quality--latency tradeoff: the first two stages produce initial actions more quickly, while the final stage adjusts their magnitudes to account for coupled network effects without changing the selected-device subset. Improvements in both AC feasibility and security-operation value suggest that this correction does more than increase perturbation strength. It helps coordinate the selected adjustments into more effective operating configurations. Uniform weighting, meanwhile, largely preserves feasibility but reduces $\bar{\mathcal J}$ through lower detectability and higher incremental cost. Although the offline actions are optimization-derived, they represent local solutions of differing quality. Uniform training reflects their prevalence, while reward weighting directs greater learning effort toward more favorable solutions, improving the expected quality of a limited candidate pool. Because these weights affect only offline training, this benefit requires no additional online computation.

\subsection{Scalability Study to Larger Power Systems}
\label{subsec:mtd_num_scalability}

We further evaluate the proposed framework on the IEEE 300-bus system, examining whether its security–operation performance and computational efficiency can be maintained as the network size and number of MTD decision variables increase.
A separate offline dataset is constructed, and the learning-based and generative models are trained for this system. We retain the scenario count, 70\%/30\% training--test split, 20\% D-FACTS coverage, $\lambda=10^{-5}$, and candidate budget $K_0=5$. All methods use the same hardware and evaluation protocol as the IEEE 118-bus study. 

\begin{table}[h]
\centering
\caption{PERFORMANCE COMPARISON OF MTD METHODS ON THE IEEE 300-bus
system. The BDD false-alarm rate is $\alpha=5\%$.}
\label{tab:mtd_num_scalability}
\footnotesize
\setlength{\tabcolsep}{4pt}
\renewcommand{\arraystretch}{1.12}
\begin{tabular}{@{}lcccccc@{}}
\toprule
Method &
\makecell{AC\\(\%)} &
$\bar{\mathcal D}_{\mathrm{FDIA}}$ &
\makecell{$\bar{\mathcal C}_{\mathrm{OP}}$\\($10^3\,\$/\mathrm{h}$)} &
$\bar{\mathcal J}$ &
\makecell{BDD\\(\%)} &
\makecell{Time\\(s)} \\
\midrule
Cost-benefit & 95.0 & 0.472 & 18.6 & 0.286 & 94.12 & 96.40 \\
RMTD         & 92.7 & 0.451 & 29.7 & 0.154 & 92.36 & 81.70 \\
DD-MTD       & 87.3 & 0.483 & 45.2 & 0.031 & 90.84 & 38.60 \\
\midrule
NS-MTD       & 72.0 & 0.501 & 37.9 & 0.122 & 91.57 & 1.86 \\
TD3-MTD      & 74.0 & 0.254 & 23.4 & 0.020 & 84.62 & 0.29 \\
\midrule
CVAE         & 85.1 & 0.412 & 30.8 & 0.104 & 74.86 & 0.64 \\
\textbf{Ours}& 91.8 & 0.782 & 5.4  & 0.728 & 97.26 & 1.43 \\
\bottomrule
\end{tabular}
\end{table}

The results are reported in Table~\ref{tab:mtd_num_scalability}. Compared to Table~\ref{tab:mtd_num_main}, the results indicate that the proposed method preserves its security-operation advantage as the system grows. Absolute detectability and cost magnitudes are compared within each system because their scales depend on network characteristics.
Meanwhile, the proposed per-action runtime increases from $0.59$ to $1.43$~s, approximately $2.4$ times, compared with an increase from $19.5$ to $96.4$~s, approximately $4.9$ times, for cost-benefit MTD. At these rates, generating five candidates sequentially would require approximately $7.15$~s, still well below the optimization-based baseline's time for one action.

\section{Conclusion}
\label{sec:conclusion}

This paper reformulates D-FACTS-enabled MTD as conditional generation of multiple reactance-perturbation alternatives. The proposed framework combines graph-conditioned diffusion models for branch selection, perturbation allocation, and continuous refinement with reward-weighted learning from diverse, AC-feasible optimization outcomes. The IEEE 118-bus results demonstrate an improved security–cost tradeoff and higher BDD detection rates than the evaluated optimization-based baselines, with substantially lower online computation. Beyond action quality, the candidate pool supports screening against device commitments and varying the deployed configuration across similar operating conditions. This provides a basis for adapting MTD deployment through candidate selection without repeatedly solving the underlying optimization problem. Future work will develop history-aware selection and candidate-refresh strategies, and evaluate their sustained effectiveness against attackers that recognize and learn recurring defense configurations.

\bibliographystyle{IEEEtran}
\bibliography{ref_MTD, ref_FDIA, ref_diffusion}

\end{document}